\documentclass[AMA,STIX1COL]{WileyNJD-v2}
\usepackage[utf8]{inputenc}
\articletype{Article Type}%

\begin{document}

\title{Wetting boundary conditions for multicomponent pseudopotential lattice Boltzmann}

\author[1,2]{Rodrigo C. V. Coelho*}

\author[1]{Catarina B. Moura}

\author[1,2]{Margarida M. Telo da Gama}

\author[1,2]{Nuno A. M. Araújo}

\authormark{Coelho \textsc{et al}}

\address[1]{Centro de Física Teórica e Computacional, Faculdade de Ciências, Universidade de Lisboa, 1749-016 Lisboa, Portugal}

\address[2]{Departamento de Física, Faculdade de Ciências, Universidade de Lisboa, P-1749-016 Lisboa, Portugal}

\corres{*Corresponding author: \email{rcvcoelho@fc.ul.pt}}


\abstract[Summary]{ The implementation of boundary conditions is among the most challenging parts of modeling fluid flow through channels and complex media.
Here, we show that the existing methods to deal with liquid-wall interactions using multicomponent Lattice Boltzmann are accurate when the wall is aligned with the main axes of the lattice but fail otherwise.  
To solve this problem, we extend a strategy previously developed for multiphase models.
As an example, we study the coalescence of two droplets on a curved surface in two dimensions. The strategy proposed here is of special relevance for binary flows in complex geometries.}


\maketitle

\section{Introduction}
\label{intro}

The lattice-Boltzmann method (LBM) is a popular technique to simulate fluid flows due to its flexibility in the implementation of boundary conditions~\cite{kruger2016lattice,succi2018lattice} (BCs). In many practical problems as in oil and food industry~\cite{Liu2015, YANG2013882, doi:10.1063/1.4921611}, a binary mixture of immiscible  liquids flows through a complex geometry where the flow dynamics is dominated by the interaction of the fluid components with the solid obstacles. Different approaches can be combined with LBM to simulate multicomponent (MC) flows, including color-gradient, free-energy and pseudopotential models~\cite{Liu2015, CHEN2014210, Liu2013, PhysRevE.92.033306}.  The latter are based on the separation of the two components according to a density dependent interparticle potential. It is generally of simple implementation and widely used~\cite{10.1002/2017wr021443, 10.1038/s41598-018-31803-w, doi:10.1063/1.5087266, PORTER20091632, doi:10.1139/cjp-2018-0126}, with possible extensions to account for more realistic equations of states, high density ratios~\cite{PhysRevE.91.023305, PENG2021110018} and frustrated coalescence between droplets~\cite{2021arXiv210106981S, montessori_lauricella_tirelli_succi_2019}.  

The interaction between the MC fluid and the solid in the pseudopotential model requires no-slip and wetting BC. The former is normally implemented by applying the bounce-back method~\cite{kruger2016lattice, KHAJEPOR2019103896}, in which the solid wall is discretized and takes a stair shape if it is curved or not aligned with the grid (off-grid), see the scheme in Fig.~\ref{drop-example-fig}A. For the wetting condition, there are many different proposals~\cite{Chibbaro2008, PhysRevE.83.046707, YU2020104638}. In the method proposed by Martys and Chen~\cite{PhysRevE.53.743}, the contact angle is controlled by the strength of the fluid-solid interaction. Chen et al.~\cite{CHEN2014210} introduced a new degree of freedom with the virtual solid density, which controls the contact angle and can be a function of the position. As the pseudopotential is a function of the density, the fluid-solid interaction is modelled in the same way as the interaction between the two components. Li et al.~\cite{PhysRevE.90.053301} proposed a scheme in which the solid density is the same as the fluid neighbor with which it is interacting. With this choice of solid density, the fluid-solid and the fluid-fluid interactions are of the same order and the spurious velocities are reduced at the fluid-solid interface, except at the contact line. Recently, Li et al.~\cite{PhysRevE.100.053313} used another choice of solid density for multiphase (MP) models, which they named improved virtual solid density scheme, based on the average density of the liquid neighbors. This scheme is able to reduce significantly the spurious velocities, including at the contact line, and the thick mass-transfer layer near the solid boundary. 

Here, we aim at investigating the implementation of wetting BCs in the MC pseudopotential model on off-grid walls. We perform a simple test consisting of a droplet initially at rest on a straight wall in the absence of any external field (e.g. gravity). When the wall is not aligned with the main axes of the lattice, the droplet slides along the wall due to asymmetric forces at the fluid-solid interface for two common wetting BCs: Martys and Chen~\cite{PhysRevE.53.743} and Li et al.~\cite{PhysRevE.90.053301}. The velocity of the sliding droplet depends on the angle between the straight wall and the axes and it becomes zero for walls aligned with the grid (on-grid), which is the test case in many works~\cite{ PhysRevE.90.053301, PhysRevE.87.053301, PhysRevE.101.043311, KHAJEPOR2019103896}. To fix this problem, we extend to binary fluids the improved virtual density scheme~\cite{PhysRevE.100.053313} and show that the droplet remain at rest when the wall is off-grid. As an application, we investigate the coalescence of two droplets on a curved surface in 2D.

This paper is organized as follows. In Sec.~\ref{sec-method}, we describe the pseudopotential MC LBM and the three BCs considered here, including the improved virtual density scheme. In Sec.~\ref{problem-sec}, we illustrate the problems caused by inappropriate choice of BCs for off-grid walls. In Sec.~\ref{coalescence-sec}, we study the coalescence of two droplets in flat and curved surfaces. In Sec.~\ref{conclusion-sec}, we close with some final remarks. 

\section{Method}
\label{sec-method}

The droplet dynamics and wetting is simulated using the pseudopotential MC model~\cite{kruger2016lattice}, which is summarized as follows. Consider two fluid components $A$ and $B$ (e.g., water and oil) that interact through a repulsive force, which is strong enough to promote demixing.

To describe the flow we use the Lattice Boltzmann method (LBM) to solve numerically the discrete Boltzmann equation, which gives the time evolution of the distribution function $f_i^{(\sigma)}$ for the component $\sigma$:
\begin{align}
  f_i^{(\sigma)}(\mathbf{x}+\mathbf{c}_i \Delta t, t+\Delta t) - f_i^{(\sigma)}(\mathbf{x}, t) = -\frac{\Delta t}{\tau^{(\sigma)}}\left[f^{(\sigma)}(\mathbf{x}, t)-f^{eq(\sigma)}(\mathbf{x}, t)\right] + S_i^{(\sigma)}(\mathbf{x}, t)\Delta t.
\end{align}
Space, with position given by $\mathbf{x}$, is discretized on a regular squared lattice. Our results are given in units such as the time step is $\Delta t = 1$ and the lattice spacing is $\Delta x = 1$. The relaxation time is related to the kinematic viscosity of the fluid: $\nu^{(\sigma)} = c_s^2 (\tau^{(\sigma)} - 1/2)$, where $c_s$ is the speed of sound given by the quadrature ($\tau^{(A)}=\tau^{(B)}$ is adopted in our simulations, meaning that both fluids have the same viscosity). We use the D2Q9 quadrature~\cite{Qian1992}, which gives the velocity discretization $\mathbf{c}_i$ (it considers the node at rest and the eight neighbors in the first belt), the speed of sound $c_s=1/\sqrt{3}$ and the discrete weights: $w_i=4/9$ for $\vert \mathbf{c}_i\vert^2=0$, $w_i=1/9$ for $\vert \mathbf{c}_i\vert^2=1$ and $w_i=1/36$ for $\vert \mathbf{c}_i\vert^2=2$. The equilibrium distribution is the expansion of the Maxwell-Boltzmann distribution up to second order in Hermite polynomials~\cite{COELHO2018144}:
\begin{align}
  {f^{eq}_{i}}^{(\sigma)} = \rho^{(\sigma)} w_i \left[ 1+ \frac{\mathbf{c}_i\cdot\mathbf{u}^{eq}}{c_s^2} + \frac{(\mathbf{c}_i\cdot{\mathbf{u}^{eq}})^2}{2c_s^4} - \frac{(\mathbf{u}^{eq})^2}{2c_s^2}  \right],
\end{align}
where the fluid velocity and density are:
\begin{align}
\mathbf{u}^{eq} = \frac{1}{\rho} \sum_\sigma \left( \sum_i f_i^{(\sigma)}\mathbf{c}_i + \frac{\mathbf{F}^{(\sigma)}\Delta t}{2}\right) , \quad \rho = \sum_\sigma \rho^{(\sigma)}.
\end{align}
$S_i^{(\sigma)}(\mathbf{x}, t)$ is the Guo forcing term~\cite{PhysRevE.65.046308}:
\begin{align}
S_i ^{(\sigma)} = w _i\left(1-\frac{\Delta t}{2\tau^{(\sigma)}} \right) \left[ \frac{\mathbf{c}_i}{c_s^2}\left(1+  \frac{\mathbf{c}_i\cdot \mathbf{u}^{eq}}{c_s^2}\right) - \frac{\mathbf{u}^{eq}}{c_s^2} \right]\cdot \mathbf{F}^{(\sigma)} .
  \label{source-eq}
\end{align}

The separation between the components is given by the Shan-Chen force~\cite{PhysRevE.47.1815,Shan1995}:
\begin{align}
\mathbf{F}^{SC (\sigma)}(\mathbf{x}) = - \psi^{(\sigma)}(\mathbf{x})\, G_{\sigma \bar \sigma} \sum _i w_i \,\psi^{(\bar \sigma)}(\mathbf{x}+\mathbf{c}_i\Delta t) \,\mathbf{c}_i \,\Delta t ,
\label{sc-force-eq}
\end{align} 
where the interaction strength is $G_{AB} = G_{BA}= 3$ throughout this paper (this leads to a repulsive force between the components strong enough to separate them) and the pseudopotential is equal to the density: $\psi^{(\sigma)} (\mathbf{x}) = \rho^{(\sigma)} (\mathbf{x})$. This force leads to the following equation of state:
\begin{align}
p =  c_s^2 \rho^{(A)} +  c_s^2\rho^{(B)} + G_{AB} c_s^2 \Delta t^2 \rho^{(A)}  \rho^{(B)},
\end{align}
where the first two terms represent the ideal gas contribution of each component and the second term describes the interaction between the two fluids. 

No-slip BCs are applied at the solid walls by means of the half-way bounce-back conditions~\cite{kruger2016lattice}. We use a switch function $\phi(\mathbf{x})$ to label the fluid nodes as zero and the solid nodes as one. To describe wetting, the sum in Eq.~\ref{sc-force-eq} runs only for fluid nodes while an adhesion force (fluid-solid interaction) is added:  
\begin{align}
\mathbf{F}^{s (\sigma)}(\mathbf{x}) = - \rho^{(\sigma)}(\mathbf{x})\, G_{\sigma s} \sum _i w_i \,S^{(\bar \sigma)}(\mathbf{x}+\mathbf{c}_i\Delta t) \,\mathbf{c}_i \,\Delta t .
\label{adhesion-eq}
\end{align}
The function $S^{(\sigma)}$ is the the pseudopotential for solid boundaries and takes different forms in the literature~\cite{KHAJEPOR2019103896, PhysRevE.100.053313, refId0, PhysRevE.83.046707}.  Martys and Chen~\cite{PhysRevE.53.743} proposed $S^{(\sigma)}(\mathbf{x})=S^{(\bar\sigma)}(\mathbf{x}) = \phi(\mathbf{x})$, with the wetting being controlled by the parameter $G_{\sigma s}$. Chen et al.~\cite{CHEN2014210}, introduced a new degree of freedom with the virtual solid density: $S^{(\sigma)}(\mathbf{x}) = \phi(\mathbf{x})\rho_s^{(\sigma)}(\mathbf{x})$, where $\rho_s^{(\sigma)}(\mathbf{x})$ is the solid density for each component $\sigma$ (may be equal for both components). In virtual solid density schemes, it is usual to set $G_{\sigma s}=G_{\sigma \bar \sigma}$ and the wetting is controlled by $\rho_s^{(\sigma)}(\mathbf{x})$. In 2014, Li, Luo, Kang and Chen~\cite{PhysRevE.90.053301} (LLKC) proposed a BC in which the solid node has the same density as the fluid node with which it is interacting: $S^{(\sigma)}(\mathbf{x}+\mathbf{c}_i\Delta t) = \phi(\mathbf{x}+\mathbf{c}_i\Delta t)\rho^{(\sigma)}(\mathbf{x})$. In this way, the fluid-solid interaction is of the same order as the fluid-fluid interaction and the wetting is controlled by $G_{\sigma s}$. In 2019, Li, Yu and Luo~\cite{PhysRevE.100.053313} (LYL) proposed an improved virtual density scheme for MP pseudopotential models able to reduce the spurious velocities at the fluid-solid interface. The authors have focused on the reduction of the thick mass-transfer layer near the solid boundary, which is another common unphysical effect when modeling wetting with MP/MC models.

Inspired by the LYL BC for MP models, we extend it to the MC pseudopotential model. The main difference is that there are now two virtual solid densities (one for each component) which, in principle, can be controlled independently in order to achieve the desided contact angle. The virtual solid density is equal to the averaged density of the fluid neighbors times a factor $\chi^{(\sigma)}$ that controls the wetting, $S^{(\sigma)}(\mathbf{x}) =  \phi(\mathbf{x})\tilde{\rho}^{(\sigma)}(\mathbf{x}) $, where:
\begin{align}
 \tilde{\rho}^{(\sigma)}(\mathbf{x}) = \chi^{(\sigma)}\frac{\sum_i w_i \rho^{(\sigma)}(\mathbf{x}+\mathbf{c}_i\Delta t) \phi(\mathbf{x}+\mathbf{c}_i\Delta t)  }{\sum_i w_i  \phi(\mathbf{x}+\mathbf{c}_i\Delta t) }.
 \label{av-dens-eq}
\end{align}
The weights and discrete velocities used to calculate the average can be those used for the Boltzmann equation (the D2Q9 in our case), but we have found that the contact angle is less sensitive on the inclination of the wall for the same $\chi^{(\sigma)}$, for higher order lattices, as will be discussed in Sec.~\ref{virtual-scheme-sec}. For the virtual solid density calculation, we use the D2Q25 lattice~\cite{PhysRevE.75.026702}, which considers all neighbors in the first and second belts:
\begin{align}
 w_i = 
\begin{cases}
    4/21, & \vert \mathbf{c}_i\vert^2=1\\
    4/45, &  \vert \mathbf{c}_i\vert^2=2\\
    1/60, &  \vert \mathbf{c}_i\vert^2=4\\
    2/315, &  \vert \mathbf{c}_i\vert^2=5\\
    1/5040, &  \vert \mathbf{c}_i\vert^2=8.
\end{cases}
\end{align}
If the parameter $\chi^{(\sigma)}$ is one, the wetting is neutral ($\theta=90^\circ$). In addition, the contact angle is more easily controlled if the value of $\chi^{(\sigma)}$ is different for the two components. Thus, we choose $\chi ^{(A)} = 1+\xi$ for the fluid that composes the droplets and $\chi ^{(B)} = 1-\xi$ for the surrounding fluid, where $\vert\xi \vert <1$. The surface is hydrophilic if $\xi<0$ and hydrophobic if $\xi>0$. Note that the virtual solid density only needs to be updated in the solid nodes at the interface (the solid nodes with at least one fluid neighbor in the first belt, see Fig.~\ref{drop-example-fig}A).

\section{Droplet on a off-grid straight wall}
\label{problem-sec}

In this section, we simulate a droplet initially at rest on an off-grid straight wall using three wetting BCs. This simple test reveals the problems with the existing schemes.

\subsection{Martys-Chen and LLKC boundary conditions}
\label{llkc-sec}

Here we test two common wetting BCs: Martys-Chen~\cite{PhysRevE.53.743} and LLKC~\cite{PhysRevE.90.053301}.

\begin{figure}[htb]
\centering{\includegraphics[width=0.8\linewidth]{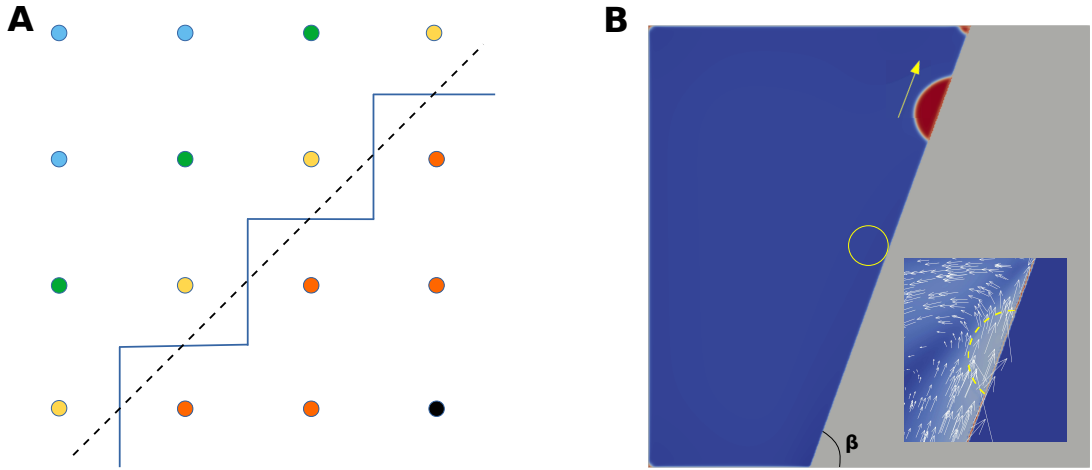}}
\vspace*{8pt}
\caption{A) Scheme of the nodes close to an off-grid straight wall (dashed line). The solid line represents the staircase approximation for the straight wall. Below the solid line are the solid nodes, which are separated in bulk-solid (black) and interface-solid (brown). Above the solid line are the fluid nodes, which can be bulk-fluid (blue), internal corners (yellow) or external corners (green). 
B) Droplet on a inclined plane for the LLKC BC with adhesion strength $G_w=-0.5$ (hydrophilic, contact angle $\theta_c \approx 63^\circ$) and inclination of the plane $\beta=70^{\circ}$ at time $t=12500$. The circle in the center represents the initial position of the droplet showing that the droplet has moved and the arrow shows the direction of the motion. The inset shows the of velocity field close to the droplet, where the arrows indicate the direction, the blue color represents velocity zero and the red represents $\vert\mathbf{u}\vert=0.032$. The yellow dashed line indicates the position of the droplet interface. The spurious velocities at the solid surface are higher than those at the interface between the two fluids and than the droplet velocity.}
\label{drop-example-fig}   
\end{figure}

We simulate a droplet on a plane with inclination $\beta$ with the horizontal axis as illustrated in Fig.~\ref{drop-example-fig}B. The solid walls are at the borders ($x=0,\, L_X-1$ and $y=0,\, L_Y-1$) and the inclined plane is given by
\begin{eqnarray*}
y= \tan \beta[x-L_X/2 - R_0 \sin\beta] + L_Y/2-R_0 \cos\beta,
\end{eqnarray*}
where $R_0$ is the initial radius of the droplet. The simulation box size is $L_X \times L_Y =400 \times 400$. It is initialized with density $\rho^{(A)}_{\text{in}}=1.0$ inside the droplet and $\rho^{(A)}_{\text{out}}=0.055$ outside for the component $A$, the main component of the droplet, and the opposite for component $B$ ($\rho^{(B)}_{\text{in}}=0.055$ and $\rho^{(B)}_{\text{out}}=1.0$). Note that the virtual solid density is not considered in these two BCs. The droplet is initially at the center with radius $R_0=18$ close to the inclined plane so as it can attach to it. We set the strength of the adhesion force to the same absolute value with different signs for the two components: $G_{A,s}=-G_{B, s} \equiv G_w$. Thus, if $G_w<0$, the solid is hydrophilic and, if $G_w>0$, it is hydrophobic. 

Figure~\ref{drop-example-fig}B shows an example of a droplet sliding upwards along the inclined plane \textit{without an external force}, due to spurious effects. In a few thousands time steps the droplet reaches the top border of the system. The small droplet at the top right corner is formed by fluid from the wetting layer which accumulates in this region where contact with the hydrophilic walls is largest. One can see in the inset that the spurious velocities are higher close to the solid than at the droplet interface. 

\begin{figure}[htb]
\centering{\includegraphics[width=1.0\linewidth]{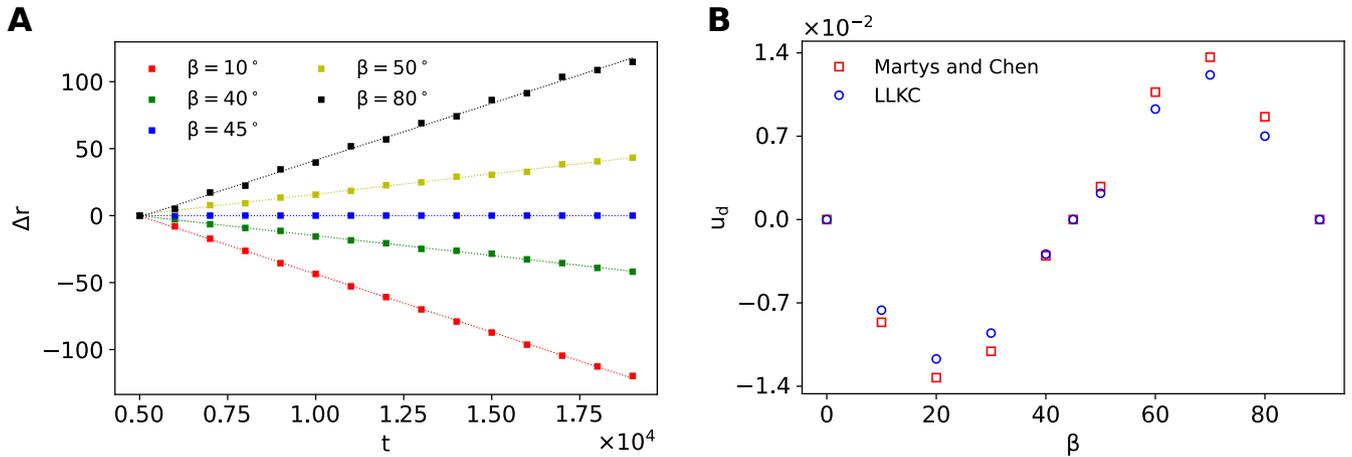}}
\vspace*{8pt}
\caption{A) Droplet displacement $\Delta r$ on an inclined off-grid plane for different inclinations $\beta$ and $G_w=0$ (neutral wetting) for the Martys-Chen BC. The squares represent the results from the simulations and the lines are linear fits to measure the droplet velocity. B) Droplet velocity $u_d$ along the plane for different inclinations $\beta$ and the two BCs: Martys-Chen and LLKC. For Martys-Chen, the adhesion strength is $G_w=-0.2$ (contact angle $\theta_c \approx 55^\circ$) and, for LLKC, the adhesion strength is $G_w=-0.5$ (contact angle $\theta_c \approx 63^\circ$).}
\label{drop-velocity-fig}   
\end{figure}

\begin{figure}[htb]
\centering{\includegraphics[width=0.6\linewidth]{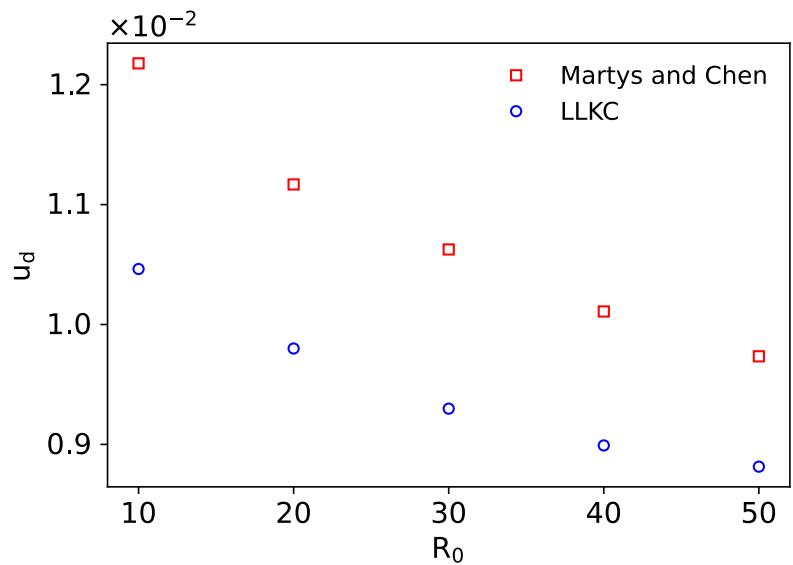}}
\vspace*{8pt}
\caption{Droplet velocity $\vert \mathbf{u}_d\vert$ as a function of the initial radius $R_0$ for the two BCs: Martys-Chen and LLKC. For Martys-Chen, the adhesion strength is $G_w=-0.2$ (contact angle $\theta_c \approx 55^\circ$) and, for LLKC, the adhesion strength is $G_w=-0.5$ (contact angle $\theta_c \approx 63^\circ$).}
\label{vel-size-fig}   
\end{figure}

Next we measure the droplet displacement along the direction of the plane, by using the position of the contact line on the right, as shown in Fig.~\ref{drop-velocity-fig}A for the Martys-Chen BC (similar results were obtained for LLKC). Most of the works in the literature (e.g.,Refs.~\cite{KHAJEPOR2019103896, PhysRevE.90.053301, PhysRevE.87.053301}) use inclination $\beta=0^\circ$ or $90^\circ$ (on-grid walls), for which the droplet remains at rest if no external force is applied. Interestingly, for $\beta = 45^\circ$, the droplet dynamics is not affected either. For all other inclinations, the droplet moves with constant velocity, of the order of $u_d\sim 10^{-2}$, until it reaches the limits of the domain. Fig.~\ref{vel-size-fig} shows that the droplet velocity decreases with the initial radius but we did not observe a static droplet. In most problems, the drop or the mixture move due to an external force or flow rate, which masks this motion due to spurious forces. 


Although we show results for a MC model, we have also observed the same behavior in the standard MP Shan-Chen model~\cite{kruger2016lattice} for these two BCs suggesting this is a general problem of pseudopotential models.


The problem of a sliding droplet in the absence of external fields is caused by asymmetric forces at the corners of the inclined plane. Because of the discretization, the off-grid wall has a stair shape at the surface, as illustrated in Fig.~\ref{drop-example-fig}A. For the D2Q9 lattice, we can separate the fluid nodes at the interface in two: internal corners, those with three solid neighbors in the first belt, and external corners, with only one solid neighbor (see Fig.~\ref{drop-example-fig}A). From Eq.~\ref{adhesion-eq}, one can see that the different number of neighbors for internal and external corners lead to forces with different magnitudes for these two type of corners. If the surface is flat as in many works, the droplet dynamics is not affected because the adhesion force has the same magnitude (except at the contact line) and points in the same direction. In the case of an inclined plane with inclination $\beta=45^\circ$, the number of the two types of corners is the same, and that is why the droplet remains at rest in this case. If the number of distinct corners is not balanced, the droplet moves because there is a net force which results in coherent flow close to the surface. We notice that using high-order lattices to calculate the forces reduces the spurious velocities in the bulk fluid as shown in Ref.~\cite{PhysRevE.73.047701}, but it does not solve the problem of unbalanced forces at the off-grid walls.

\subsection{Improved virtual density scheme}
\label{virtual-scheme-sec}

\begin{figure}[htb]
\centering{\includegraphics[width=0.6\linewidth]{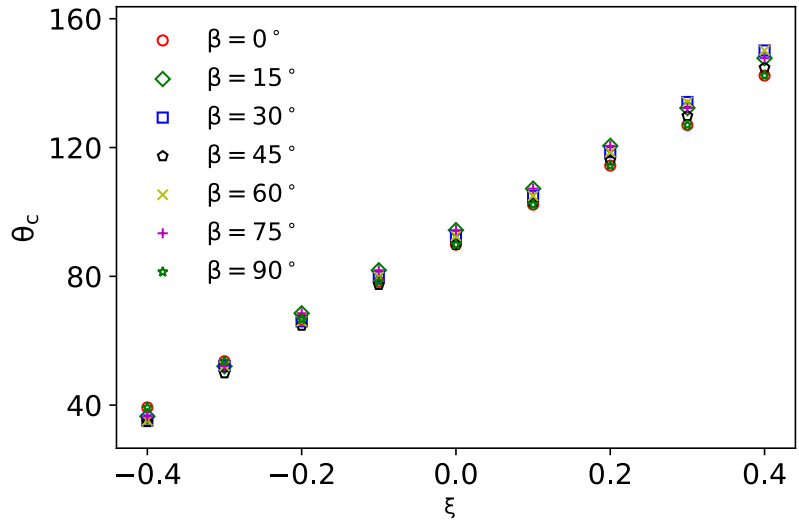}}
\vspace*{8pt}
\caption{Contact angle as a function of the parameter $\xi$ for different inclinations $\beta$ of the plane using the improved virtual density scheme.}
\label{contact-angle-fig}   
\end{figure}
Now we repeat the simulation of a droplet on an off-grid plane with the improved virtual density scheme (similar to the LYL BC). The system dimensions are $L_X \times L_Y = 1000 \times 1000$, $\tau=1.88$, $R_0=100$ and the same initial and BCs as in Sec.~\ref{llkc-sec}. In these BCs, we identify the solid nodes at the interface as the solid nodes with at least one fluid neighbor (see Fig.~\ref{drop-example-fig}), which is done only once. The virtual solid density at the interface nodes is updated for the two components at every time step according to Eq.~\eqref{av-dens-eq}. 

With the improved virtual density scheme, we observe that the droplet remains static for any inclination $\beta$ and contact angle $\theta_c$. We verify that, as in the multiphase model~\cite{PhysRevE.100.053313}, the spurious velocities are significantly reduced close to the solid surface, specially at the contact line, when compared to the Martys-Chen and LLKC schemes. We then fit a circle using the interface between the two fluids of the static droplet (with density $\rho^{(A)}=0.5$) and calculate the angle between the tangent to the circle at the contact line and the plane. Fig.~\ref{contact-angle-fig} shows the result for different values of $\xi$ (parameter
which controls the wetting, see Sec.~\ref{sec-method}) and inclinations of the plane. 
The variation of the contact angle for the same $\xi$ and different $\beta$ is small and decreases when we use the D2Q25 lattice to calculate the virtual solid density instead of the D2Q9 lattice. This results from the higher isotropy of the D2Q25 and the fact that it considers more fluid nodes to calculate the virtual solid density in Eq.~\ref{av-dens-eq}. Thus, the virtual solid density becomes more homogeneous along the off-grid walls and less dependent on the wall inclination. Notice for instance that, with the D2Q9 lattice, the internal corners in off-grid walls consider only one fluid node to calculate the virtual solid density, while the external corners consider three.
The measured angle may also vary slightly with the resolution due to the diffuse interface and with the fluid viscosity as lower viscosities induce higher spurious velocities. The last problem can be reduced with the usual methods such as the multi-relaxation time collision operator and higher order lattices~\cite{kruger2016lattice, PENG2019104257, PhysRevE.73.047701}.  
Since this BC solves the issue of unbalanced forces at the fluid-solid interface, we can use it to study problems involving wetting on off-grid walls.

\section{Coalescence on flat and curved surfaces}
\label{coalescence-sec}

The coalescence of droplets has been studied experimentally and theoretically, mainly in 3D and for two liquid droplets in a gas with negligible density~\cite{Paulsen6857, PhysRevLett.106.114501, PhysRevLett.95.164503, Xia23467, PhysRevE.75.056315, doi:10.1063/1.4803178, sprittles_shikhmurzaev_2014}. The coalescence is driven by surface tension when the two droplets touch while the resistance arises from the viscosity and inertia. The time evolution of the bridge radius $b$ (minimum distance between the two interfaces of the bridge or between the surface and the interface) often results in two power laws: $b\sim t$ in the initial viscous regime and $b \sim t^{\frac{1}{2}}$ in the inertial regime. Similarly to the coalescence of single component droplets (neglecting the gas), results for MC systems reveal that $b \sim t^{\frac{1}{2}}$ in the inertial regime~\cite{doi:10.1063/1.4729791, GAC2011355}. In 2D, the analytical calculations predict $b\sim t^{0.857}$ in the viscous regime for single component droplets~\cite{doi:10.1111/j.1151-2916.1984.tb19692.x}, which was later observed in a numerical study~\cite{sprittles_shikhmurzaev_2014}. 

\begin{figure}[htb]
\centering{\includegraphics[width=1.0\linewidth]{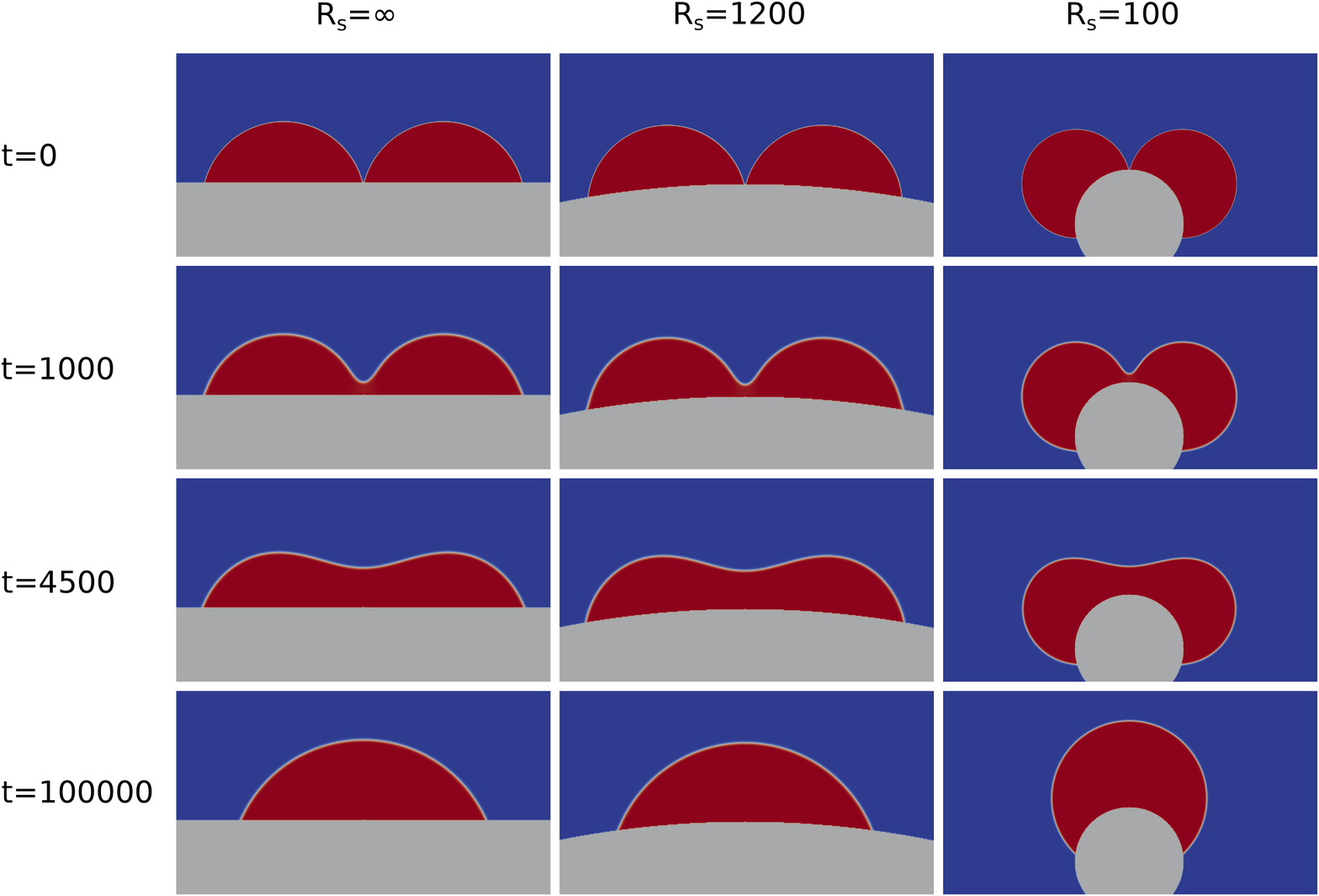}}
\vspace*{8pt}
\caption{Snapshots of the coalescence of two droplets on circular surfaces with different radii $R_s$ at different time steps for $Oh = 0.286$. The initial droplet radius is $R_0=100$ and the kinematic viscosity $\nu=0.46$. }
\label{samples-fig}   
\end{figure}

\begin{figure}[htb]
\centering{\includegraphics[width=1.0\linewidth]{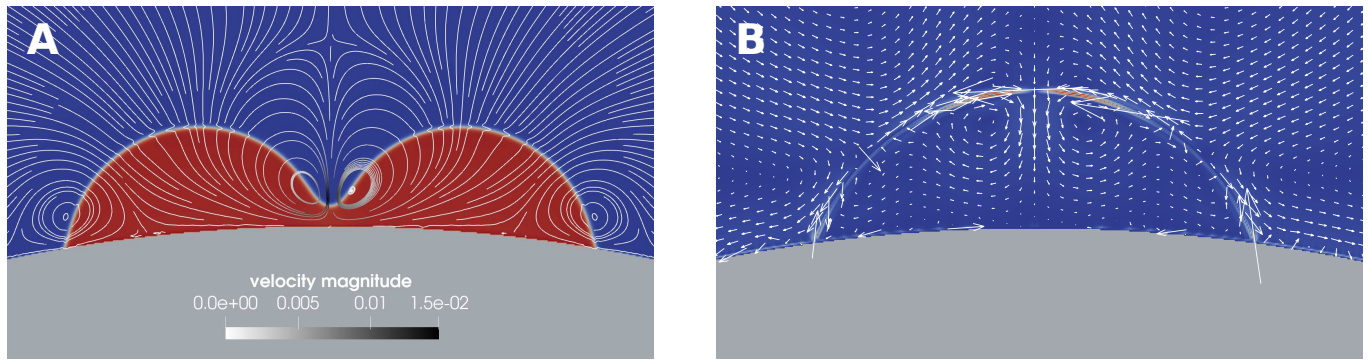}}
\vspace*{8pt}
\caption{A) Streamlines of the flow field during the coalescence for $R_s=1200$ and $t=1000$. The colors of the streamlines indicate the magnitude of the velocity field as indicated by the colorbar. B) Velocity field for $R_s=1200$ and $t=100000$ showing the spurious velocities, where the maximum velocity is $\vert \mathbf{u} \vert=0.0018$. }
\label{streamlines-fig}  

\end{figure}

In this section, we consider the coalescence of two droplets immersed in a different liquid with the same density and viscosity. The droplets at rest have a contact angle $\theta_c=71^\circ$ ($\xi=-0.2$) on a surface, which may be circular with radius $R_s$ or flat ($R_s=\infty$). The simulation box size is $L_X \times L_Y = 1400 \times 1000$ (except in the convergence study). The two droplets, with initial radius $R_0$,  are initialized close to each other and make an angle $\theta_c$ with the surface. The curved surface is given by the equation $(x-L_X/2)^2 + (y-(L_Y/2-R_s))^2 = R_s^2$ and the two droplets have smoothed initial interfaces: the density field of the fluid that composes the droplets $\rho_A$ and the surrounding fluid $\rho_B$ are, respectively:
\begin{align*}
 &\rho_A =  \rho_2+\frac{1}{2}(\rho_1-\rho_2)\left[1 - \tanh\left( \frac{(x-x_{c1})^2+(y-y_c)^2 - R_0^2}{\delta^2} \right)\right] + \frac{1}{2}(\rho_1-\rho_2)\left[1 - \tanh\left( \frac{(x-x_{c2})^2+(y-y_c)^2 - R_0^2}{\delta^2} \right)\right]\\
 & \rho_B = \rho_1-\frac{1}{2}(\rho_1-\rho_2)\left[1 - \tanh\left( \frac{(x-x_{c1})^2+(y-y_c)^2 - R_0^2}{\delta^2} \right)\right] - \frac{1}{2}(\rho_1-\rho_2)\left[1 - \tanh\left( \frac{(x-x_{c2})^2+(y-y_c)^2 - R_0^2}{\delta^2} \right)\right],
\end{align*}
where $\rho_1=1$, $\rho_2=0.055$, $\delta=8$, $x_{c1}=L_X/2-R_0 \sin(\theta_c)-1$, $x_{c2}=L_X/2+R_0 \sin(\theta_c)+1$ and $y_c=L_Y/2-R_0 \cos(\theta_c)$. Fig.~\ref{samples-fig} shows the coalescence for three different surface radii and Fig.\ref{streamlines-fig}A shows the streamlines for one particular case. We notice that two vortices in each droplet form during the coalescence and that the velocity is higher close to the liquid bridge. When the coalescence is complete, we measured the spurious velocities, which were found to be one order of magnitude smaller than the physical velocities and are significantly reduced close to the solid surface when compared to the LLKC BC (see Fig.~\ref{drop-example-fig}).

\begin{figure}[htb]
\centering{\includegraphics[width=0.6\linewidth]{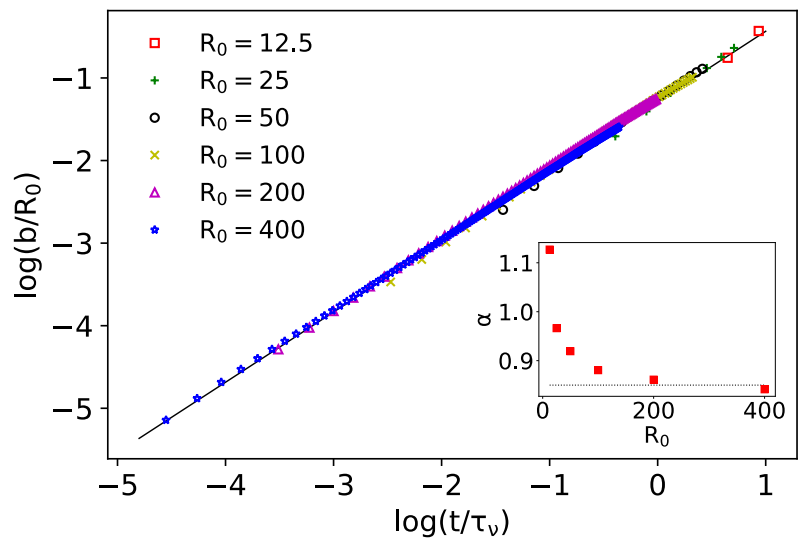}}
\vspace*{8pt}
\caption{ Slope $\alpha$ of the linear fit to data in the viscous regime for different resolutions represented by the initial radius of the droplets $R_0$ on a flat surface. The Ohnesorge number is kept constant at $Oh = 0.286$ while all the dimensions scale with $R_0$. The solid line is the reference value $\alpha_{\text{ref}}=0.857$. The inset shows the measured slope for each resolution. The horizontal line is $\alpha_{\text{ref}}$.}
\label{convergence-fig}   
\end{figure}

First, we obtained the time dependence of the bridge radius $b$ on a flat surface in the viscous regime. A relevant quantity to characterize the droplet coalescence is the Ohnesorge number~\cite{Paulsen6857}: $Oh = \eta /\sqrt{ \rho \sigma R_0}$, where $\eta=\rho \nu$ is the shear viscosity and $\sigma$ is the surface tension. For $G_{AB}=3$, we obtain $\sigma =0.026$ through the Laplace test~\cite{kruger2016lattice}. We keep the same Ohnesorge number as in Fig.~\ref{samples-fig}, $Oh = 0.286$, for different resolutions to test the convergence. The smallest system size used is $175 \times 125$ (with initial droplet radius $R_0=12.5$ and $\nu = 0.16$) and the largest $5600 \times 4000$ (with $R_0=400$ and $\nu=0.92$). In Fig.~\ref{convergence-fig} we plot the time evolution of the bridge radius in reduced units, where the time is divided by the viscous time scale $\tau_\nu = \eta R_0 / \sigma$. The slope from the simulations converges to the analytical prediction, $b\sim t^{0.857}$, for 2D coalescence (as shown in the inset).

\begin{figure}[htb]
\centering{\includegraphics[width=1.0\linewidth]{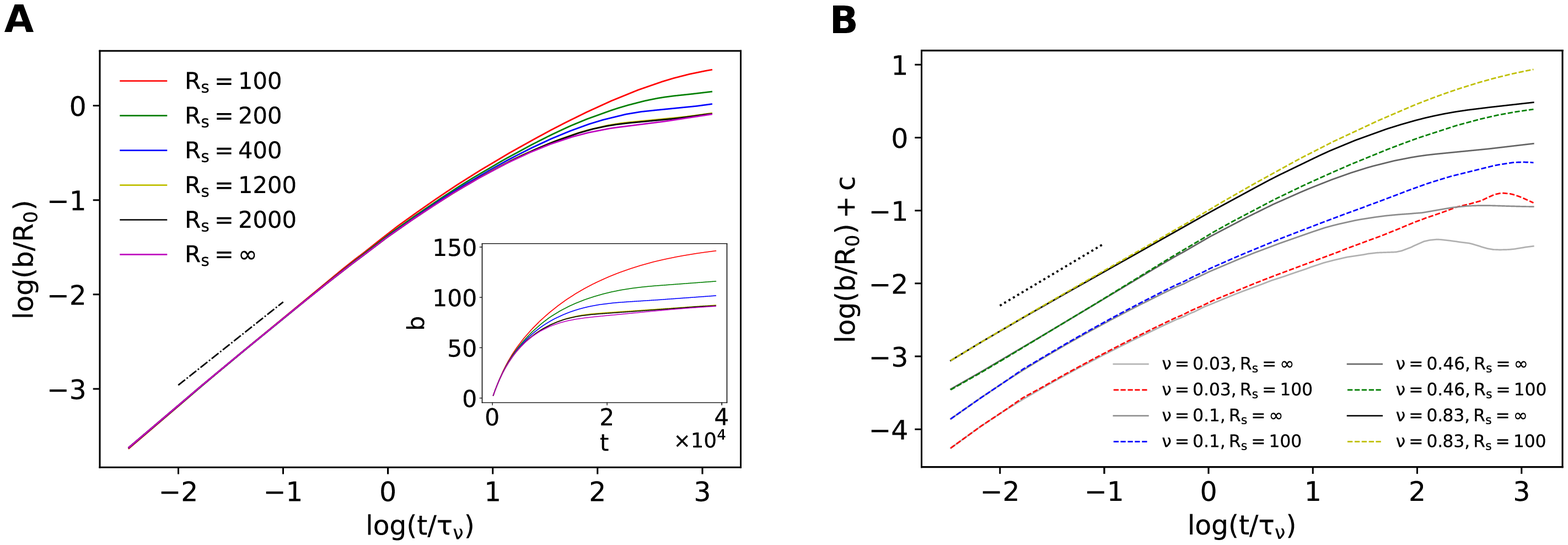}}
\vspace*{8pt}
\caption{A) Time evolution of the bridge radii $b$ for different surface radius $R_s$ and $Oh = 0.286$. The inset shows the raw data in lattice units. B) Time evolution of the bridge radius $b$ for different viscosities $\nu$ and two surface radii: $R_s=\infty$ (flat) and $R_s=100$. The curves are shifted by a constant $c$ for each viscosity so as to distinguish them at the beginning (only $\nu=0.46$ is in the original place). The doted line indicates the reference slope $\alpha_{\text{ref}}=0.857$.}
\label{curved-fig}   
\end{figure}

Next, we changed the curvature of the surface over which the droplets coalesce. The result in Fig.~\ref{samples-fig} suggest that the curvature does not affect the coalescence in the first steps. This is confirmed in Fig.~\ref{curved-fig}A, where we plot the time evolution of the bridge radius for surfaces with different radii, including the flat case for comparison. In the viscous regime, the curvature of the surface does not change the time evolution $b\sim t^{0.857}$ observed for a flat interface. The difference becomes evident when the bridge radius becomes larger than $R_0/3$, in the cross-over between the viscous and inertial regimes. This difference occurs because the total amount of fluid A depends on the surface radius $R_s$ due to the choice of the initial conditions. From the inset of Fig.~\ref{curved-fig}A we see that the curves converge to different final radii. It was observed in experiments that the crossover between viscous and inertial regimes occurs at earlier times for smaller viscosities~\cite{PhysRevLett.95.164503}. In Fig.~\ref{curved-fig}B we show the evolution of the bridge radius for different viscosities and two curvatures ($R_s=\infty$ and $R_s=100$). We note that the initial power law ($\sim t^{0.857}$) is followed for longer times when the viscosity is higher, as expected. In addition, we observe oscillations in the inertial regime for smaller viscosities.


\section{Conclusion}
\label{conclusion-sec}

In this work, we have pointed out that widely used wetting BCs result in unbalanced forces close to off-grid solid walls, which significantly affects the dynamics of the system in multicomponent pseudopotential lattice Boltzmann methods. To illustrate the point, we initialize a droplet at rest on a straight off-grid plane and observed that it moves in the absence of external fields, when the Martys-Chen and Li-Luo-Kang-Chen BCs are used. The droplet velocity depends on the angle that the off-grid plane makes with the axes, being zero when the plane is aligned to the axes or when the angle is $45^\circ$. This becomes relevant in simulations in complex geometries such as in porous media since most of the walls are off-grid and the spurious velocities caused by unbalanced forces at the surface contaminates the results. Although we have shown results for a multicomponent model, we have also observed the same problem of a sliding droplet in an off-grid wall in the Shan-Chen multiphase model~\cite{kruger2016lattice}, indicating that this is a generic issue of the pseudopotential models.

We then extended to binary fluids a recent virtual solid density scheme~\cite{PhysRevE.100.053313} originally proposed and tested for the multiphase model. This approach fixes the problem of a sliding droplet on an off-grid wall (due to unbalanced forces close to the solid). Similar problems have been fixed in other multicomponent models. Refs.~\cite{10.1002/2017WR020373, 10.1002/fld.4226, AKAI201856} propose boundary conditions for wetting in curved surfaces using the color-gradient model and Ref.~\cite{VANDERSMAN20132751} does the same for the free-energy model.

As an example, we studied the coalescence of two droplets on a curved surface in 2D. We showed that the time evolution of the bridge radius follows the expected power law in the viscous regime for a flat surface and that this evolution is unchanged by the surface curvature. Differences due to the surface curvature appear in the inertial regimes at latter times.

\section*{Acknowledgments}

We acknowledge financial support from the Portuguese Foundation for Science and Technology (FCT) under the contracts: PTDC/FIS-MAC/28146/2017 (LISBOA-01-0145-FEDER-028146), UIDB/00618/2020 and UIDP/00618/2020.

\bibliographystyle{plain}
\bibliography{references}

\end{document}